\documentclass[%
 reprint,
superscriptaddress,
 amsmath,amssymb,
prl
]{revtex4-1}

\usepackage{graphicx}
\usepackage{subfigure}
\usepackage{dcolumn}
\usepackage{bm}
\usepackage{color}%
\usepackage{amsmath}
\usepackage{verbatim}
\usepackage{mathtools}
\usepackage{epstopdf}
\usepackage{float}

\def\beq{\begin{equation}}
\def\eeq{\end{equation}}

\begin{document}


\title{
Sub-anomalous diffusion and unusual velocity distribution evolution \\
in cooling granular gases: theory}

   \author{Raphael Blumenfeld}
   \email{rbb11@cam.ac.uk}
\affiliation{Gonville \& Caius College, Cambridge University, Trinity St., Cambridge CB2 1TA, UK}

\date{\today}

\begin{abstract}

There is no agreement in the  literature on the rate of diffusion of a particle in a cooling granular gas. Predictions and model assumptions range from the conventional to very exotic dependence of the mean square distance (MSD) on time. This problem is addressed here by calculating the MSD from first-principles. The calculation is based on random-walking and it circumvents the common use of continuum equations and equations of states, which involve approximations that erode at low gas particle densities.
The MSD is found to increase logarithmically with time -- slower than even in anomalous diffusion. This result is consistent with the well-established Haff's law for the decay of the kinetic energy, which is also derived along the way from the same first principles. This derivation also pins down Haff's time constant, alleviating the usual need for a fitting parameter.
The diffusion theory is then used to calculate explicitly the time evolution of any initial particle velocity distribution, yielding an unusual functional form. 
The limitations of the theory are discussed and extensions to it are outlined.

\end{abstract}



\maketitle

\noindent{\bf Introduction}

Granular matter in general and granular gases in particular are ubiquitous in nature and are relevant to many technologies and natural phenomena. 
Granular gases consist of isolated particles that are affected negligibly by thermal fluctuations and dissipate kinetic energy on collisions. In the absence of external injection of energy, the collisional dissipation reduces the overall initial kinetic energy, which `cools' the gas, effecting ever slowing diffusion of particles, and gives rise to clusters formation \cite{GoZa93,Mc93,BrDu97,BrPo00}. 
There are conflicting claims in the literature regarding the diffusive behaviour of particles in these dissipative systems. (i) The diffusion is conventional , i.e., the mean squared distance (MSD) proportional to time, $\langle R^2\rangle = Dt$, with a suppressed diffusion coefficient $D$ \cite{BrDu97,DuBr02,BrPo05}. (ii) It is conventional, but with the coefficient $D$ depending on time in a complex manner \cite{BrPo00}. (iii) It is anomalous, $\langle R^2\rangle \sim t^{\alpha}$, with $0<\alpha<1$ \cite{BrRu15}. (iv) It has a complicated time dependence unless reflective boundary conditions are introduced, which reduce it to the conventional form~\cite{Heetal01}. (v) The MSD has a more complex dependence on time, with different behaviour in different regimes~\cite{Meetal14,Boetal16}. It has been recognised that the main difficulty may be in the erosion of basic assumptions when extending continuum equations, be those hydrodynamics and/or equations of state, to low particle densities~\cite{CuSt79,Haff83,MiDi,SaBr99}. 
This state of affairs reflects the absence of a first-principles theory of diffusion in cooling granular gases, which starts from the particle-scale kinetics and upscaled to the continuum.  \\

This is the first aim here -- to derive an explicit expression from first principles for the time dependence of the MSD of a diffusing particle. The derived relation is parameter-free, depending only on the gas density, the restitution coefficient, and the initial particle velocity distribution. The diffusion is found to be `sub-anomalous', in the sense that it increases {\it logarithmically} with time, which is even slower than in anomalous diffusion. This derivation reproduces the continuum-based Haff's law~\cite{Haff83} showing that the two are consistent. The present derivation circumvents the limiting assumptions of the traditional continuum-based derivation of Haff's law and, consequently, requires no free parameter.\\
The second aim, also achieved here, is to use the new diffusion theory to determine explicitly the evolution of the particles velocity distribution. 
The obtained results are consistent with existing numerical observations and some extensions are outlined. \\

\noindent{\bf Particle diffusion in a freely cooling gas}:
Consider a gas of particles in three dimensions, assumed, for simplicity, to be of similar sizes and similar masses $m$. All collisions are presumed to involve only two particles at a time and to be inelastic, with each colliding particle losing a constant fraction, $\epsilon$, of its momentum. An extension of the following to collisions in which $\epsilon$ depends weakly on velocity~\cite{Raetal99,Gretal09}, such as for iron particles, is outlined in the concluding discussion. The energy dissipation may be by radiation or by exciting intra-particle degrees of freedom, which decay eventually also by radiation. Between collisions, the particles move ballistically. The initial particles number density, $\rho$, is assumed spatially uniform and the velocity distribution isotropic. The initial particle velocities are sufficiently high such that the following analysis is restricted to the time before clusters form.  \\

Taking a mean field approach, a particle suffers a collision every time it travels a mean free path, $l_0=\gamma\rho^{-1/3}$, with $\gamma=(3/4\pi)^{1/3}\Gamma(4/3)\approx0.55396$~\cite{He1909}. 
Crucially, since the motion between collisions is ballistic, the statistics of the spatial trajectory of a diffusing particle are the same as in a conventional random walk. The only difference is that the momentum loss stretches the time spent between two successive collisions by $1/\epsilon$. 
The MSD of such $N$-steps three-dimensional random walkers, starting at velocity $v_0$, is then $\langle R^2(N)\rangle = 3l_0^2 N$ and the time to make the $N$ steps is 
\begin{equation}
t = \frac{l_0}{v_0}\sum_{k=1}^N \epsilon^{1-k} = \frac{\epsilon^{1-N}-\epsilon}{1-\epsilon}\ \frac{l_0}{v_0} \ .
\label{Time}
\end{equation}
From (\ref{Time}), the velocity after the $N$th step is
\begin{equation}
v_N=v_0\epsilon^N= \frac{v_0}{1+t/\tau_0} \ ,
\label{Speed}
\end{equation}
with $\tau_0\equiv \epsilon l_0/[(1-\epsilon) v_0]$. Using (\ref{Speed}), the MSD is 
\begin{equation}
\langle R^2\rangle = 3l_0^2 N = \frac{3\ln{\left(1+t/\tau_0\right)}}{\ln{(1/\epsilon)}}l_0^2  \ .
\label{MSD1}
\end{equation}
From (\ref{Time}) and (\ref{Speed}), the mean kinetic energy per unit volume is
\begin{equation}
E_k = \frac{m\rho v_0^2}{2\left(1+t/\tau_0\right)^2}  \ .
\label{KE}
\end{equation}
This decay follows the frequently observed Haff's law \cite{Haff83}, supporting the derivation of (\ref{MSD1}). However, Haff's law was derived originally from continuum equations with viscosity-based energy dissipation. The present derivation via the particle-scale random-walk formalism has several advantages. 1) Being based on particle-scale physics, it conveniently circumvents the assumptions involved in the continuum approach, which become questionable at low gas density. 2) It provides the explicit dependence of $\tau_0$, known as the Haff time scale, on $v_0$, $\rho$ and $\epsilon$, which are readily measurable. As such, this obviates the need for the fitting parameter, used in Haff's derivation, and makes possible direct tests of the above results in real and numerical experiments. 3) It makes possible an explicit calculation of the evolution of the particle velocity distribution, as shown below. 
Significantly, eq. (\ref{MSD1}) shows that the MSD increases {\it logarithmically} with time, in contrast to several models in the literature~\cite{BrDu97,DuBr02,BrPo05,BrPo00,BrRu15}. 
The logarithmic dependence makes this diffusion `sub-anomalous', i.e., slower than even the very slow anomalous diffusion, $\langle R^2\rangle\sim t^{\alpha(<1)}$\cite{Shetal93,MeKl00,Meetal14}. 
A weak dependence of $\epsilon$ on $v$, such as found in \cite{Gretal09}, would slow down further the diffusion rate, as discussed in the concluding discussion. \\

\begin{figure}[htbp]
   \centering
   \includegraphics[width=.4\textwidth]{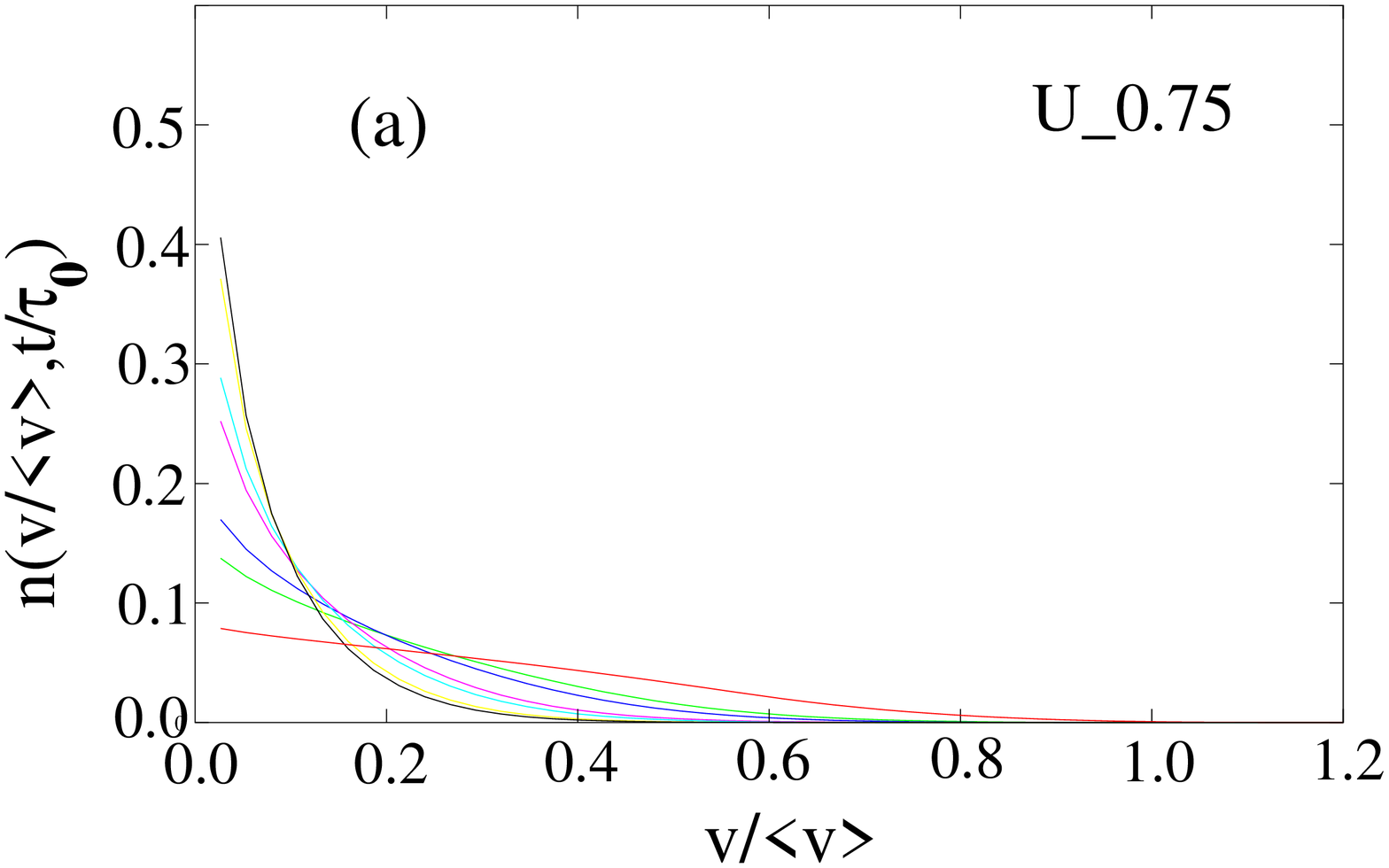}
   \includegraphics[width=.4\textwidth]{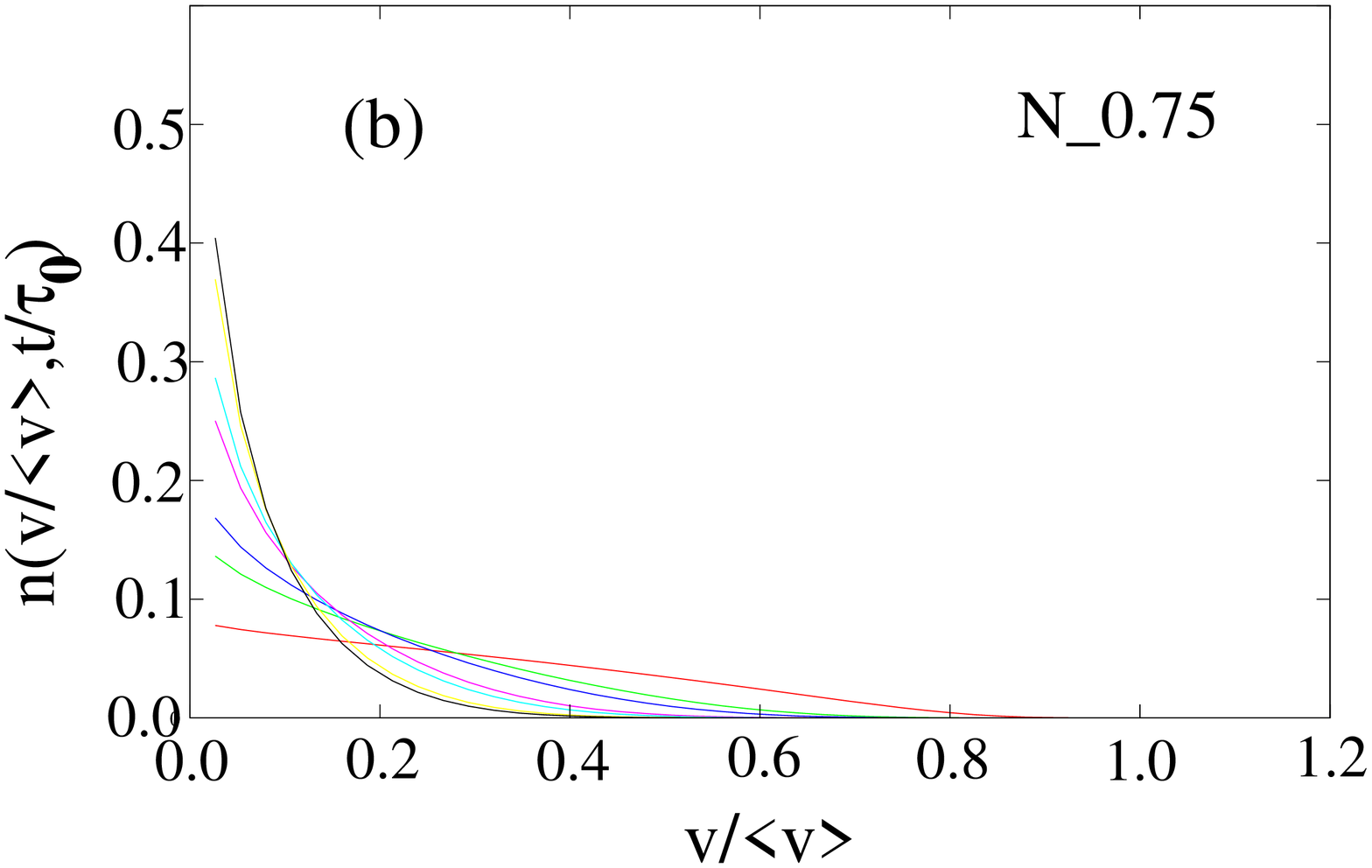}
   \caption{A generic evolution of the unconditional velocity PDF for an initial uniform PDF (a) and normal PDF (b), shown at times: $t/\tau_0 = 0.1$ (red), $0.5$ (green), $1.0$ (blue), $5.0$ (purple), $10.0$ (light blue), $50.0$ (yellow), $100.0$ (black). The inter-particle restitution coefficient is $0.75$. The differences between these two cases in the long-time limit $t/\tau_0\to\infty$, as well as compared with restitution coefficients $0.50$ and $0.90$, are hardly noticeable.}
   \label{fig:Pv}
\end{figure}

\noindent{\bf The velocity distribution}:
The above results are used next to determine the evolution of the velocity distribution. 
The probability that a particle, starting at speed $v_0$ at $t=0$, covers a distance $R$ after experiencing $N$ collisions is $P_N(R)=Ae^{-3R^2/Nl_0^2}$, with $A=1/\int_{0}^{\infty} P_N(R) dR$. 
After $N$ collisions, and before any cluster nucleates, the particle's speed is $v_N=v_0\epsilon^N$. 
Since $N$ is a distributed variable, the probability that $v_N$ is between $v$ and $v+dv$ at time $t>0$ is 
\begin{equation}
p(v,t\!\mid\! v_0) dv = \sum_{N=1}^\infty\! P_N(N,R)\ \delta\!\left( N-\frac{\ln{\left(v_0/v\right)}}{\ln{\left( 1/\epsilon\right)}}\right)\! dv 
\label{PDFv1}
\end{equation}
which yields, after substituting for $P_N$,
\begin{equation}
n(v,t\!\mid\! v_0) \equiv \rho p(v,t\!\mid\! v_0) = C\rho\left(1+t/\tau_0\right)^{1/\ln{\left(v/v_0\right)}} ,
\label{PDFv2}
\end{equation}
with $0<v< v_0$ and the normalisation factor is
\begin{equation}
C = \frac{1}{2v_0\sqrt{\ln{\left(1+t/\tau_0\right)}}\mid\! K_1\left(2\sqrt{\ln{\left(1+t/\tau_0\right)}}\right)\!\mid} \ ,
\label{NormC}
\end{equation}
where $K_\nu(x)$ the modified Bessel function of the second kind \cite{GrRy}. 
Integrating (\ref{PDFv2}) over all possible initial speeds, $v_0$, distributed as $n_0\left(v_0\right)$, the later-time distribution is 
\begin{equation}
n(v,t) = \int\limits_{v}^{v_{max}} C \left(1+t/\tau_0\right)^{1/\ln{\left(v/v_0\right)}} n_0(v_0) dv_0 \ ,
\label{PDFv3}
\end{equation}
with $v_{max}<\infty$ the highest possible speed at $t=0$. 
With $C$ dependent on $v_0$, this integral is difficult to calculate analytically, but it was evaluated numerically for three values of the restitution coefficient, $\epsilon=0.50, 0.75$, and $0.90$, and for two very different initial distributions, $n_0\left(v_0\right)$, one uniform and the other normal. Typical examples are shown in Fig. \ref{fig:Pv} for $\langle v_0\rangle=30$ (arbitrary units), $\epsilon=0.75$ and $\rho = 1$ (arbitrary units). Both distribution converge to an almost identical distribution $n(v,t)$. 
It should be noted that relation (\ref{MSD1}) and the distribution of $N$ are valid for sufficiently many collisions and only the distributions for $t/\tau_0 \gtrapprox {\cal{O}}(10)$ are practically relevant. The numerical calculations then show that memory of the initial velocity distribution is lost within $t = \approx {\cal{O}}(10)\times\tau_0$. \\
For the specific case of an initial gas, whose all particles have the same speed $u_0$, (\ref{PDFv3}) reduces to
\begin{equation}
n(v,t) = C(u_0,t) \rho \left[1+ \frac{(1-\epsilon)u_0}{l_0}t\right]^{1/\ln{\left(v/u_0\right)}}  \ .
\label{PDFv4}
\end{equation}
Otherwise, by the mean value theorem, relation  (\ref{PDFv4}) is a good approximation with $u_0$ some value satisfying $0<v<u_0<v_{max}$. \\

\noindent{\bf Conclusion and discussion} 

To conclude, a theory of the particles diffusion in free-cooling granular gases has been developed from first-principles. The loss of momentum on collisions was shown to give rise to {\it sub-anomalous} diffusion, with the MSD increasing logarithmically with time. This increase is slower than even the very slow anomalous diffusion, in which $\langle R^2\rangle\sim t^{-\alpha}$, with $0<\alpha<1$. This result, which is free of any fitting parameter and depends only on the gas density, the restitution coefficient and the initial particle velocity, calls into question studies based on the assumption of conventional diffusion. 
It is further supported by the consistency with  the well-established Haff's law \cite{Haff83}, which the present derivation recovers also without any fitting parameter. 
Moreover, reproducing Haff's law in this approach fixes the Haff time $\tau_0$, therefore improving on the original analysis, which includes a free parameter. This approach also avoids the problem of the erosion of the continuum assumptions at low particle density~\cite{Haff83,MiDi,FoPo08}. 

The sub-anomalous diffusion behaviour was then used to derive explicitly the evolution of the particles velocity distribution, eq. (\ref{PDFv2}). The distribution was found  to evolve into an unusual form -- an algebraic power of an inverse logarithm of the velocity -- and it was shown to shift to low velocities within $t/\tau_0\sim \mathcal{O}(10)$. Furthermore, initially uniform and normal distributions of the same mean were shown to converge to an almost identical form, suggesting that the velocity distribution loses memory of its initial form also within a similar time scale.

These results agree quantitatively well with existing numerical and experimental reports~\cite{Kuetal97,MiLu04,Lu05,Isetal08,Taetal09,Haetal18,Waetal18} on the decay of the mean velocity (eq. (\ref{Speed})) and the drop in the kinetic energy (eq. (\ref{KE})). The evolution of the speed distribution, however, agrees with existing such studies only qualitatively. This is most likely because the unusual functional form found here, eqs. \ref{PDFv3} and (\ref{PDFv4}), was not known in those reports and more conventional fits were used. Moreover, since the velocity distribution converges into a form very different from a Gaussian, analyses in terms of deviations from a Gaussian form were unlikely to reveal it.

It is worth discussing the range of validity of the theory. The analysis is valid for the initial cooling stage before large clusters form. Specifically, a cluster nucleates when the relative velocity of two colliding particles is below a certain threshold, $v_c$, which leaves them with insufficient momenta to bounce away. This is known as the bouncing barrier~\cite{BouncBar}. 
To quantify this condition, suppose the colliding particles, each of mass $m$ and size $d$, experience only a mutual gravitational attraction. The kinetic energy required to bounce away must overcome the gravitational potential, namely,
\begin{equation}
\frac{G m^2}{D} \lesssim \frac{mv_c^2}{2} \quad \Rightarrow \quad v_c\gtrsim  \sqrt{2Gm/D} \ ,
\label{Gravity}
\end{equation}
with $G=6.67408\times10^{-11}$m$^3$/(kg s$^2$) the gravitational constant and $D=\mathcal{O}(d)$ is their separation.
Using (\ref{PDFv3}), the probability that they form a two-particle nuclear cluster is 
\begin{equation}
p_c(t) = \int\limits_{0}^{v_{max}} \frac{n(v_1,t)n(v_2,t)}{\rho^2}H\left[v_c - \!\mid\! v_1\!-\! v_2\!\mid\right]\!dv_1\!dv_2 \ ,
\label{BB1}
\end{equation}
with $H$ the Heavyside step function. 
$p_c(t)$ increases with time as $n(v,t)$ shifts toward low speeds. 
The effective medium approach is valid for initial times somewhat beyond $p_c\ll1$ and until clusters grow sufficiently large to effect large fluctuations of the spatial density. This happens when the typical size of the growing clusters starts approaching the typical distance between them. 

Another limitation, made to simplify the calculations, is the assumption of a constant restitution coefficients. The analysis can be extended to coefficients that decrease weakly with relative collision velocity~\cite{Gretal09}, e.g., $\epsilon(v) = \epsilon_0[1-\phi f(v)]$, with $0<\phi\ll 1$ and $f(v)$ an arbitrary analytic function. In this case, the speed after $N$ collisions can be approximated as 
\begin{equation}
v_N = v_0\epsilon^N\left[1 - \phi G(v_0) - \mathcal{O}(\phi^2\epsilon)\right]  \ ,
\label{BB1}
\end{equation}
with $G=\sum_{k=1}^{N-1} f\left(\epsilon^kv_0\right)$. This leads to a higher momentum loss and stretches further the time between collisions, which slows down the increase of the MSD with time beyond relation (\ref{MSD1}) and pushes the diffusion deeper into the sub-anomalous regime. 
However, how this affects the gas cooling rate is unclear: while the faster loss of momenta on collisions accelerates the cooling rate, the longer time intervals between collisions slows it down. 
It would be interesting to study quantitatively the competition between these opposing effects. \\

\noindent{\bf Acknowledgments}:
I thank Profs. M.-Y. Hou and K. Huang for introducing me to the problem and for initial discussions. The hospitality of the Cavendish Laboratory is gratefully acknowledged.\\

\end{document}